\pdfoutput=1

\documentclass[11pt,twoside,a4paper,cmspaper,final,collab]{cms-tdr}
\def\svnVersion{176995}\def\cmsCernNo{2012-311}\def\cmsCernDate{\today}\def\cmsMessage{Submitted to the European Physical Journal C}

\begin{document}\cmsNoteHeader{SMP-12-015}

\hyphenation{had-ron-i-za-tion}
\hyphenation{cal-or-i-me-ter}
\hyphenation{de-vices}

\RCS$Revision: 162149 $
\RCS$HeadURL: svn+ssh://svn.cern.ch/reps/tdr2/papers/SMP-12-015/trunk/SMP-12-015.tex $
\RCS$Id: SMP-12-015.tex 162149 2012-12-19 22:05:15Z ilyao $
\newlength\cmsFigWidth
\ifthenelse{\boolean{cms@external}}{\setlength\cmsFigWidth{0.85\columnwidth}}{\setlength\cmsFigWidth{0.4\textwidth}}
\ifthenelse{\boolean{cms@external}}{\providecommand{\cmsLeft}{top}}{\providecommand{\cmsLeft}{left}}
\ifthenelse{\boolean{cms@external}}{\providecommand{\cmsRight}{bottom}}{\providecommand{\cmsRight}{right}}
\newcommand{\mjj}{\ensuremath{m_{jj}}\xspace}
\providecommand{\lum}{\ensuremath{\,\text{(lum.)}}\xspace}
\providecommand{\hyph}{-\penalty0\hskip0pt\relax} 
\newcommand{\mt}{\ensuremath{m_{\mathrm{T}}}\xspace}

\newlength{\columnfigure}
\newlength{\tripleColFig}
\ifthenelse{\not{\boolean{cms@external}}}{%
  \setlength{\columnfigure}{0.65\columnwidth}
  \setlength{\tripleColFig}{0.48\columnwidth}
}
{%
  \setlength{\columnfigure}{0.9\columnwidth}
  \setlength{\tripleColFig}{0.68\columnwidth}
}
\cmsNoteHeader{SMP-12-015} 
\title{Measurement of the sum of WW and WZ production with W+dijet events in pp collisions at $\sqrt{s} = 7$\TeV}

\date{\today}

\abstract{
A measurement of the inclusive WW+WZ diboson production
cross section in proton-proton collisions is
reported, based on events containing a leptonically decaying
W boson and exactly two jets. The data sample, collected at $\sqrt{s} = 7$\TeV with the CMS detector at the LHC,
corresponds to an integrated luminosity of 5.0\fbinv.
The measured value of the sum of the inclusive WW and WZ cross sections is
$\sigma(\Pp\Pp\to\PW\PW+\PW\cPZ) = 68.9 \pm 8.7\stat\pm 9.7\syst \pm 1.5\lum\unit{pb}$,
consistent with the standard model prediction of $65.6 \pm 2.2$\unit{pb}.
This is the first measurement of WW+WZ production in pp collisions using
this signature.
No evidence for anomalous triple gauge couplings is found
and upper limits are set on their magnitudes.
}

\hypersetup{%
pdfauthor={CMS Collaboration},%
pdftitle={Measurement of the sum of WW and WZ production with W+dijet events in pp collisions at sqrt(s) = 7 TeV},%
pdfsubject={CMS},%
pdfkeywords={CMS, physics}}

\maketitle 

The gauge symmetry of the standard model (SM) fixes the
triple gauge boson couplings that determine the self-interactions
of W and Z bosons. The pair production of vector gauge bosons
allows a direct test of the electroweak sector of the SM~\cite{PhysRevD.48.2182}. Observation
of anomalous triple gauge boson couplings would be an indication of
physics beyond the SM.

We report the first measurement of WW+WZ diboson
production in pp collisions in the semileptonic final state at
the Large Hadron Collider (LHC), where
one W boson decays leptonically ($\ell\nu$, with $\ell=\Pe,\mu$) while the other boson
(W or Z) decays hadronically ($jj$), giving rise to two energetic jets in
the final state.
Previous measurements in this channel at the Tevatron p$\bar{\rm p}$
collider include the recent CDF~\cite{CDF-diboson-lvjj-2010} and
D0~\cite{Abazov:2011cb, Abazov:2012} results.
The advantage of reconstructing WW+WZ in the $\ell\nu jj$ decay mode
over the purely leptonic final states~\cite{Chatrchyan:2011tz,:2012ks,ATLAS:2012wz,ATLAS:2012ww} is the larger
branching fraction of W and Z bosons to quarks. This
advantage is partially offset by the larger backgrounds
in the $\ell\nu jj$ channel, coming mainly from W+jets production.
In contrast to the fully leptonic decay of WW pairs, the semileptonic
process permits a direct measurement of the boson transverse
momentum (\pt). The sensitivity of WW production to the WW$\gamma$ coupling
and of WW and WZ production at high boson transverse momentum to the WWZ coupling
makes these processes particularly useful as a probe of anomalous triple gauge boson couplings.

The data correspond to an integrated luminosity of $5.0 \pm
0.1\fbinv$ collected in 2010 and 2011 with the Compact Muon Solenoid
(CMS) detector in pp collisions at $\sqrt{s} = 7\TeV$ at the CERN
LHC. The CMS experiment~\cite{CMS:2010} uses a right-handed coordinate system, with
the origin at the nominal interaction point, the $x$ axis pointing to
the center of the LHC ring, the $y$ axis pointing up, perpendicular to the
plane of the LHC ring, and the $z$ axis along the counterclockwise beam direction. The
polar angle $\theta$ is measured from the positive $z$ axis and the
azimuthal angle $\phi$ is measured in the $x$-$y$ plane. The
pseudorapidity is defined as $\eta = -\ln[\tan(\theta/2)]$. The
central feature of the CMS apparatus is a superconducting solenoid of
6\unit{m} internal diameter, providing a magnetic field of
3.8\unit{T}. Within the field volume are silicon pixel and strip trackers and several calorimeters.
The tracking system covers the range $|\eta| < 2.5$
and provides a track momentum
resolution of 1\% at 100\GeV. The lead tungstate crystal
electromagnetic calorimeter (ECAL) covers $|\eta| < 3$ with an energy
resolution of about $3\%/\sqrt{E}$, where $E$ is in
\GeVns~\cite{CMS-PAS-EGM-10-003}.
The brass/scintillator hadron calorimeter (HCAL) covers $|\eta| < 3.0$ with
an energy resolution of $100\%/\sqrt{E}$.
The muon system consists of gas-ionization detectors inside and around the steel return yoke, and is capable of reconstructing and identifying muons within
$|\eta| < 2.4$. Extensive forward calorimetry complements the coverage
provided by the barrel and endcap detectors. The CMS detector is nearly
hermetic, allowing for measurements of the missing transverse energy
($\MET$) in the event.  A two-tier trigger system selects the events
of interest.

The data were collected with a suite of single-lepton triggers mostly
using \pt thresholds of 24\GeV for muons
and 25--32\GeV for electrons.
To preferentially select events with on-shell W bosons,
the single-electron triggers also require minimum
thresholds on $\MET$ in the range 0--25\GeV and on the
transverse mass \mt of the electron plus $\MET$ system in the range 0--50\GeV.
The overall trigger efficiency is about 94\% (90\%) for muon (electron) data,
with a small dependence (a few percent) on \pt and $\eta$.
Simulated events are corrected for the trigger efficiency as
a function of lepton \pt and $\eta$, and in the case
of electrons also as a function of $\MET$. Simulated events are used to
develop and validate the methods used in the analysis.

The {\MADGRAPH}5~1.3.30~\cite{madgraph5} event generator produces
parton-level events with a \PW\ boson and up to four partons on the
basis of matrix-element (ME) calculations.  The ME--parton shower (ME--PS)
matching scale $\mu$ is taken to be 20\GeV~\cite{Hoche:2006ph}, and
the factorization and renormalization scales are both set to $q^2 =
M_{\PW}^2 + p_{\mathrm{T},\PW}^2$.  Samples of \ttbar\ and Drell--Yan
events are also generated with \MADGRAPH.  Single-top production is
modeled with \POWHEG~1.0~\cite{POWHEG}.  Multijet and diboson samples
(\PW\PW, \PW\cPZ, \cPZ\cPZ) are generated with
\PYTHIA~6.422~\cite{Sjostrand:2006za}.  \PYTHIA provides the parton
shower simulation in all cases, with parameters of the underlying
event set to the Z2 tune~\cite{PythiaTuneZ2}.  The set of parton
distribution functions used is \textsc{cteq6ll}~\cite{CTEQ}.
A \GEANTfour-based simulation~\cite{GEANT4} of the CMS
detector is used in the production of all Monte Carlo (MC)
samples. Multiple proton-proton interactions within a bunch crossing (pileup)
are simulated, and the triggers are emulated. All simulated
events are reconstructed and analyzed as measured collision events.

Events are selected with one well-identified and isolated lepton (muon
or electron), large missing transverse energy, and exactly two high-\pt jets.
Muons are reconstructed within $|\eta|<2.1$ with the inner tracker and the muon system~\cite{MUONPAS}.
Electrons are reconstructed  from tracks in the tracker
pointing to energy clusters in the ECAL, within $|\eta|<2.5$,
excluding the transition region between the barrel and endcap,
$1.44<|\eta|<1.57$~\cite{CMS-PAS-EGM-10-004}.
Muons (electrons) are required to have \pt
greater than 25\GeV (35\GeV).
The lepton candidates are required to be consistent with having originated from the primary
vertex of the event, which is chosen to be the vertex with the highest
$\sum \pt^2$ of its associated tracks. According to the simulation,
this requirement provides the correct assignment for the primary
vertex in more than 99\% of the cases
in both signal and background events.
Charged leptons from
W boson decays are expected to
be isolated from other activity in the event.
The sum of transverse momentum or energy in the tracker, ECAL, and HCAL,
within a surrounding cone
of $\Delta R \equiv \sqrt{(\Delta\eta)^2+(\Delta\phi)^2} <0.3$,
excluding the lepton itself, is required to be
less than 10\% of the measured
$\pt$ of the muon, or less than 5\% of the measured $\pt$ of the electron.
Here $\Delta\eta$ and
$\Delta\phi$ are the differences in pseudorapidity
and in azimuthal angle, respectively.
To reduce the backgrounds from fully leptonic decays, such as Drell--Yan and electroweak diboson
processes, we exclude events in which there is any other
loosely identified lepton (with $\pt>10$\GeV for
muons and $\pt > 20$\GeV for electrons) in the event.

Jets are reconstructed from calorimeter and tracker information using
a particle-flow technique that combines
information from several subdetectors~\cite{CMS-PAS-PFT-09-001}.  The anti-\kt
clustering algorithm~\cite{antikt,fastjetmanual} with a distance
parameter of 0.5 is used.  Jets that overlap with isolated leptons
within $\Delta R=0.3$ are not considered.  Jet-energy corrections are
applied to account for the nonlinear energy response of the calorimeters
and for other instrumental effects~\cite{jetpas}.
These corrections are based on in situ measurements using dijet,
$\gamma$+jet, and Z+jet data samples~\cite{Chatrchyan:2011ds}.
Pileup collisions and the underlying event add to the energy of the
reconstructed jets.  The median energy density from pileup is
evaluated in each event and the corresponding energy is subtracted
from each jet~\cite{fastjet1}.  In addition, charged tracks
that do not originate from the primary vertex are not considered for jet
clustering~\cite{Cacciari:2008gn}.
We verified that these procedures
sucessfully remove the dependence of jet response on the number of
interactions in a single event.
A jet-quality requirement,
primarily based on the energy balance between charged and neutral
hadrons in a jet, is applied to remove poorly reconstructed jets. Only
events containing exactly two jets with $\pt>35$\GeV and within $|\eta|<2.4$
are selected for the analysis.  To reduce contamination from \ttbar
background, events are discarded if one or more jets pass high-efficiency
b-quark jet identification criteria
based on the presence of a secondary vertex within the jet~\cite{btag}. An
accurate $\MET$ measurement is essential to distinguish the W
signal from multijet backgrounds and to reconstruct the full event
kinematics of the WW system.  We use $\MET$ measured in the event
using the full particle-flow reconstruction~\cite{Chatrchyan:2011tn}
and require $\MET>25$ (30)\GeV for the muon (electron) channel.
To reduce the background from processes that do not contain
$\textnormal{W}\to\ell\nu$ decays, we require that the transverse
mass of the W candidate exceed 30\GeV (50\GeV)
in muon (electron) data~\cite{VBTF}.

\begin{figure*}[tbh]
\includegraphics[width=\tripleColFig]{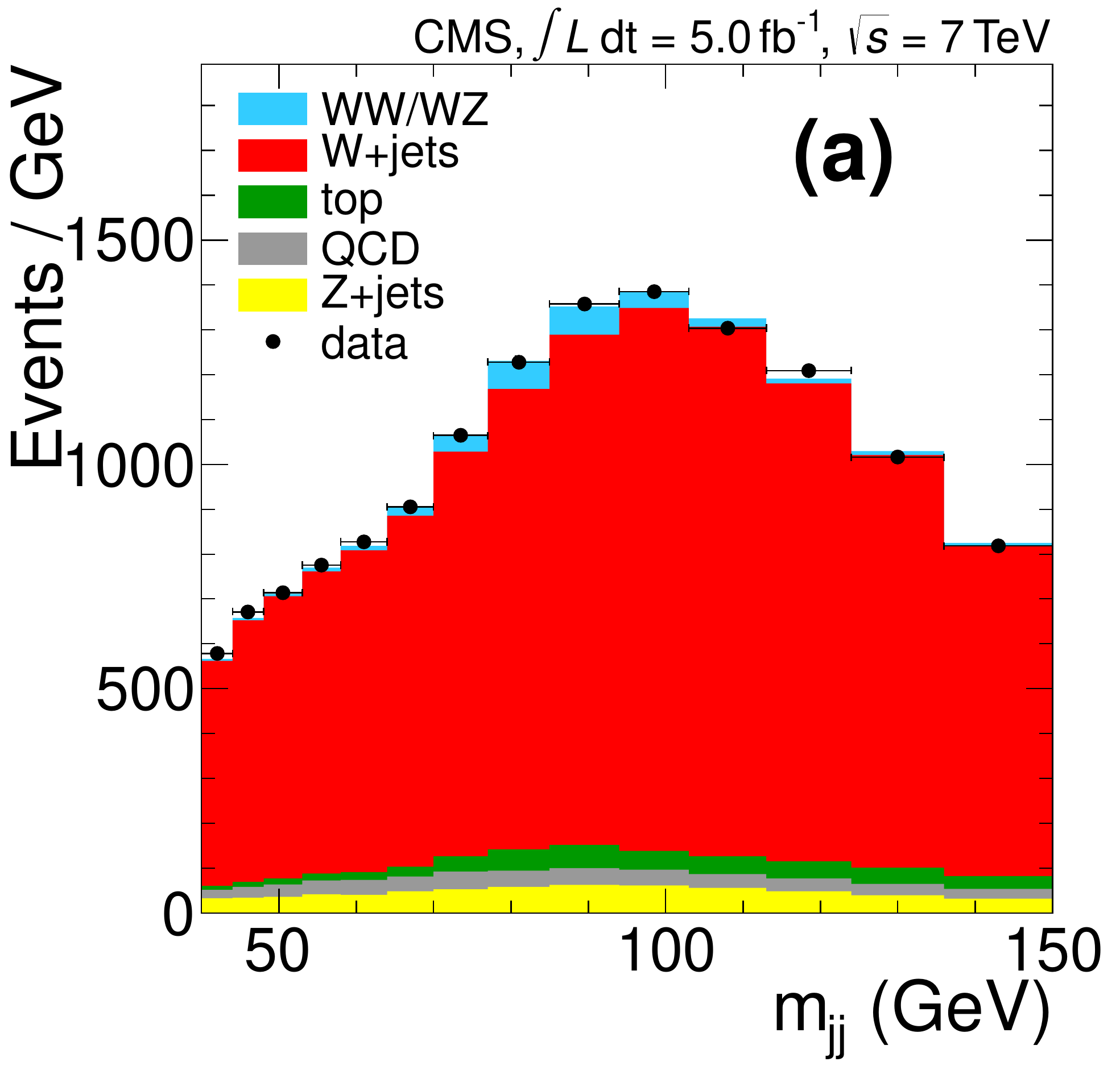}
\includegraphics[width=\tripleColFig]{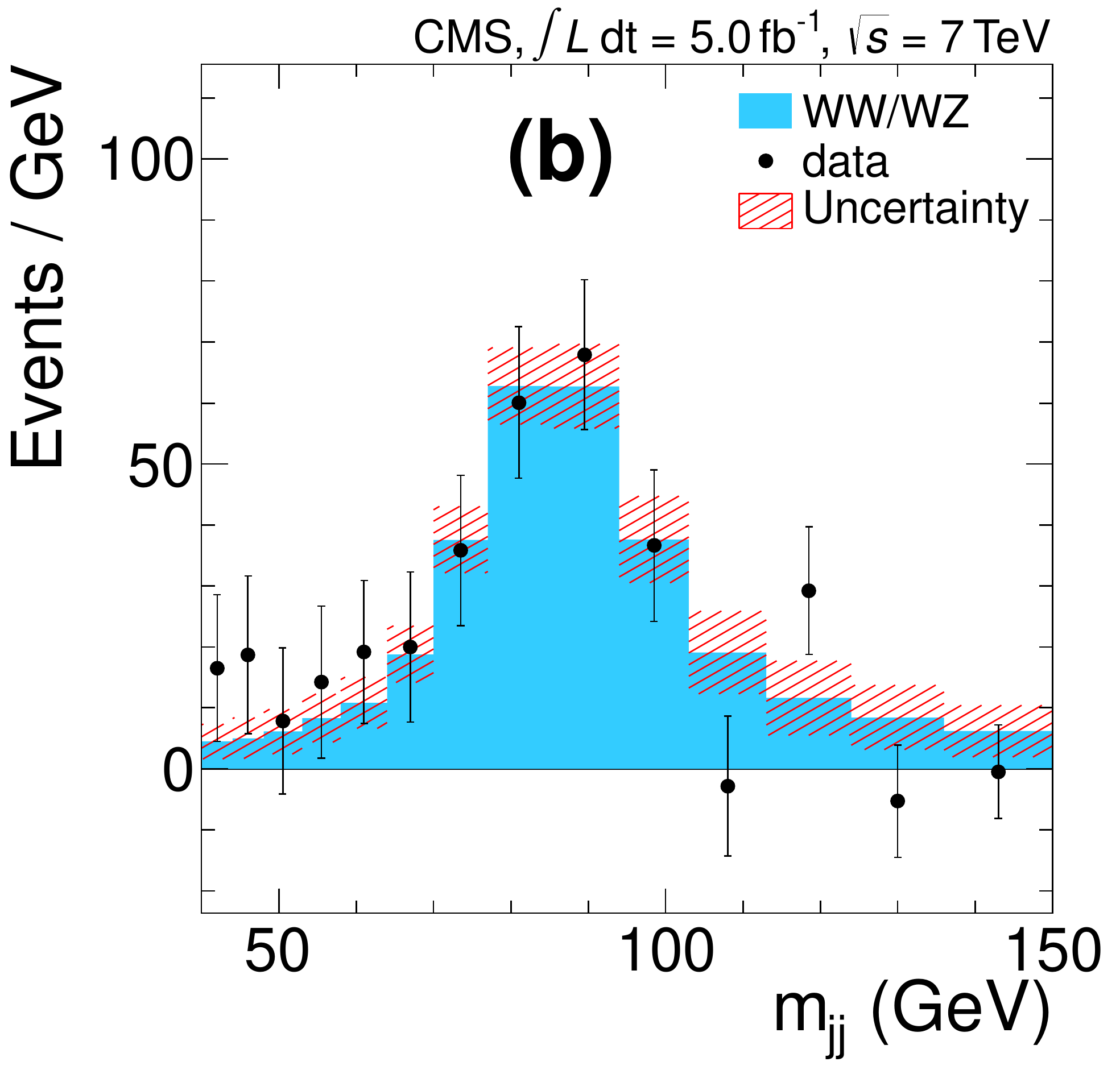}
\includegraphics[width=\tripleColFig]{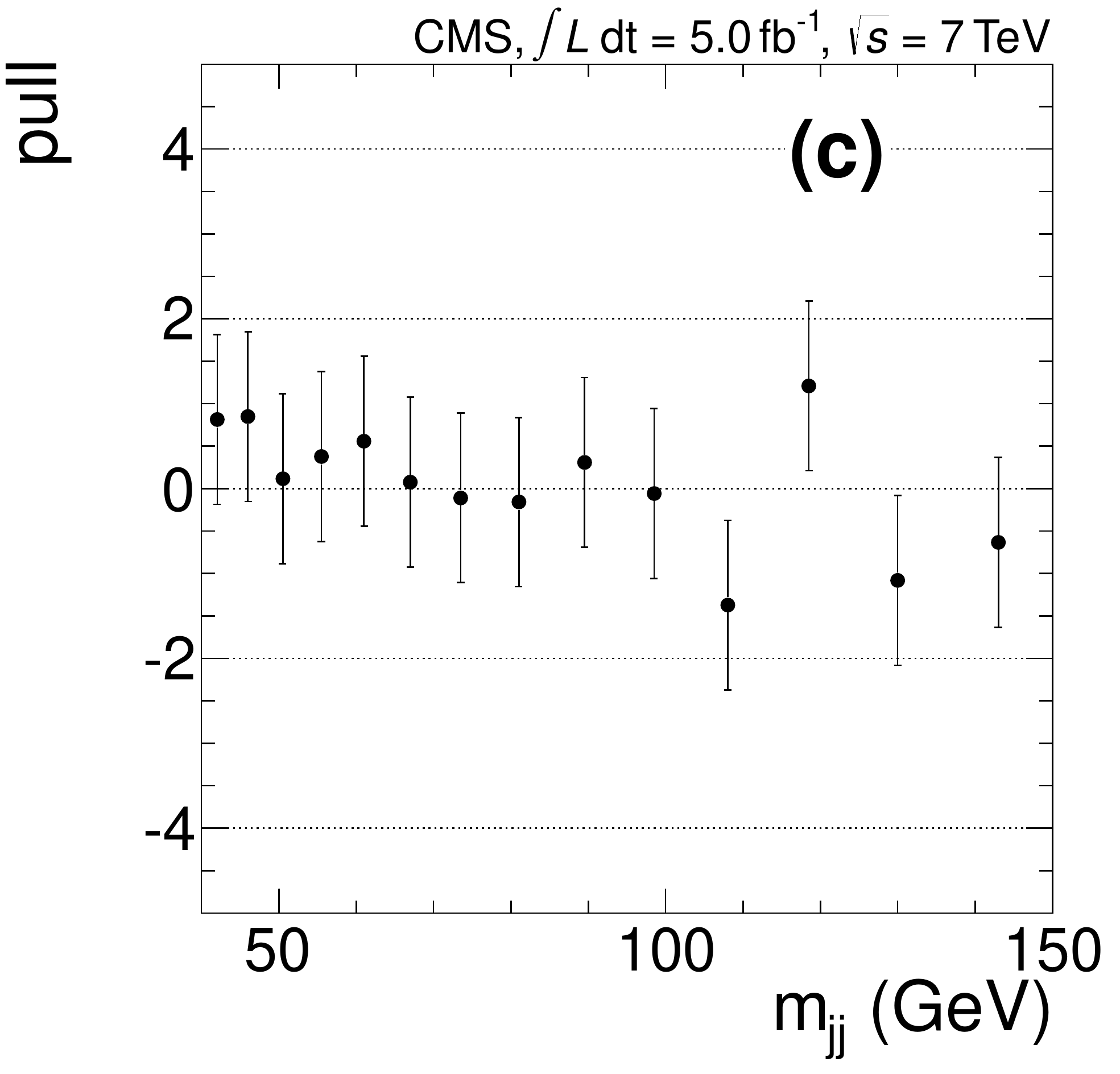}
\caption{(a) Distribution of the dijet invariant mass in data,
with the binning chosen based on the resolution and fit projections of the relevant components overlaid.
(b) The dijet invariant mass after subtraction of all components except the
electroweak WW+WZ processes.  The error bars represent the
statistical uncertainties and the hatched bands represent the systematic
uncertainties.  (c) The normalized
residual or pull: $(\text{data} - \text{fit})/(\text{fit~uncertainty})$.
}
\label{fig:Fig1}
\end{figure*}

We measure the dijet mass (\mjj) distribution, as shown in Figure~\ref{fig:Fig1}(a).
The relative contributions of the known SM processes are determined using an unbinned maximum-likelihood fit over
the mass range 40--150\GeV.  The fit is performed separately
for the muon and electron channels since their background compositions differ.
\begin{table}[bt]
\centering
\caption{Treatment of background \mjj shapes and normalizations in a fit
to the data.
The cross section values are calculated with the programs cited on the
corresponding rows.
The background normalizations are constrained to Gaussian
distributions with the listed central values and widths. The treatment of
multijet events is described in the text.\label{tab:Table0}}
\begin{tabular} {lcl}
\hline
   Process             &    Shape      & Constraint on normalization \\
   \hline
   Diboson (WW+WZ)     &    MC  & Unconstrained \\
   W+jets              &    MC  & $31.3\unit{nb} \pm 5\%$ (NLO) \cite{MCFM}\\
   \ttbar\             &    MC  & $163\unit{pb} \pm 7\%$ (NLO) \cite{Kidonakis:2010dk}\\
   Single top          &    MC  & $85\unit{pb} \pm 5\%$(NNLL) \cite{Kidonakis:2010tc,Kidonakis:2011wy,Kidonakis:2010ux}\\
   Drell--Yan+jets     &    MC  & $3.05\unit{nb}\pm 4.3\%$ (NNLO) \cite{FEWZ}\\
   Multijet (QCD)      &    data  & \MET fit in data \\
   \hline
 \end{tabular}
\end{table}
Table~\ref{tab:Table0} lists the SM processes included in the fit.
The normalization of the diboson WW+WZ contribution is a free parameter.
The normalizations of the background components
are allowed to vary within Gaussian constraints around their central
values. For multijet events, this central value is obtained from an
independent two-component fit to the $\MET$ distribution
which determines the corresponding fraction in the data~\cite{VBTF}.
The fit uncertainty is used as a constraint on the multijet contribution.
The central values for all other processes are obtained from
next-to-leading-order (NLO) or higher-order calculations, and
the constraints are taken from the theoretical uncertainties listed 
in Table~\ref{tab:Table0}. With the
exception of multijet production, the shape of the \mjj distribution
for all processes is obtained from simulation.  Multijet
events contribute to the total background when jets are misidentified as isolated
leptons. Their \mjj shape can be derived from data events with
lepton candidates that fail the isolation requirements. The fluctuations 
in the shapes and yields of subleading backgrounds have a minor impact on the overall fit.

The \mjj spectrum of the dominant W+jets component is
described using the shape from \MADGRAPH simulation after taking into account
the uncertainties due to the factorization and renormalization scale
(both equal to $q$) and ME--PS matching scale $\mu$~\cite{Alwall:2007fs}:
\begin{linenomath}
\begin{align}
F_{\text{W+jets}} = \alpha\, & \mathcal{F}_{\text{W+jets}} (\mu_{0}^2, q'^2) +
\beta\, \mathcal{F}_{\text{W+jets}} (\mu'^2, q_{0}^2) \nonumber \\
&+ (1-\alpha-\beta)\, \mathcal{F}_{\text{W+jets}} (\mu_{0}^2, q_{0}^2)\,,
\label{eqWpjetsShape}
\end{align}
\end{linenomath}
where $\mathcal{F}_{\text{W+jets}}$ denotes the \mjj shape from
simulation.  The parameters $\mu_0$ ($\mu'$) and $q_0$ ($q'$)
correspond to the default (alternative) values of $\mu$ and $q$,
respectively. The parameters $\alpha$ and $\beta$ are
free to vary during the fit and remain within the physical ranges ($0 \leq \alpha,\beta \leq 1$ and $1-\alpha-\beta\geq 0$).  We take $\mu' = 2 \mu_0$ or $0.5 \mu_0$
($q' = 2 q_0$ or $0.5 q_0$), depending on which alternative sample
provides a better fit to the data.
Thus, the fit probes variations of a factor of two in both $\mu$ and $q$ 
(with the corresponding shape fluctuations accounted for when setting exclusion limits).
\begin{table*}[htbp]
  \topcaption{Event yields determined from a maximum-likelihood fit to the data.
  The total uncertainty is computed using the full covariance matrix. Owing to a higher kinematic threshold the
  product of acceptance $\times$ efficiency is smaller for the electron
  channel. The term $\mathcal{A}\varepsilon$ includes W and Z branching fractions~\cite{Beringer:1900zz}.
  \label{tab:yields}}
\centering
\begin{tabular} {lll}
\hline
        Process          &    Muon channel                  & Electron channel \\
\hline
        Diboson (WW+WZ) &   1900 \,\,\,$\pm$ 370            & 800 \,\,\,\,\,\,$\pm$ 310   \\
        W plus jets     &   67380 $\pm$ 590		    & 31640 $\pm$ 850  \\
        $\ttbar$        &   1660 \,\,\,$\pm$ 120	    & 950 \,\,\,\,\,\,$\pm$ 70   \\
        Single top      &   650 \,\,\,\,\,\,$\pm$ 30        & 310 \,\,\,\,\,\,$\pm$  20   \\
        Drell--Yan+jets &   3610 \,\,\,$\pm$ 160	    & 1410 \,\,\,$\pm$ 60   \\
        Multijet  (QCD) &   300 \,\,\,\,\,\,$\pm$ 320       & 4190 \,\,\,$\pm$ 870   \\
\hline
    Data                &    75419           &  39365           \\
    Fit $\chi^2/N_{\text{dof}}$ (probability) &   9.73/12 (0.64)     &  5.30/12 (0.95)  \\
\hline
    Acceptance $\times$ efficiency ($\mathcal{A}\varepsilon$) & $(5.15 \pm 0.24) \times 10^{-3}$ & $(2.63 \pm 0.12) \times 10^{-3}$ \\
\hline
    Expected WW+WZ yield from simulation & 1700 $\pm$ 60 & 870 $\pm$ 30  \\
\hline
  \end{tabular}
\end{table*}

Figure~\ref{fig:Fig1}(a) shows the observed \mjj distribution for
both channels combined, together with the fitted projections of the
contribution of various SM processes.  Figure~\ref{fig:Fig1}(b)
shows the same distribution after subtracting all SM contributions
from data except for WW+WZ events.
Figure~\ref{fig:Fig1}(c) shows the pull distribution,
i.e. the normalized residual defined as $(\text{data} -
\text{fit})/(\text{fit~uncertainty})$, where the fit uncertainty
is computed at each data point by propagating the
uncertainty in the normalization coefficients.
The yields of various SM components, as determined by the fit,
are reported in Table~\ref{tab:yields}.

In order to ensure robustness against fit parameters and to account for corresponding biases 
we validate the fit procedure by performing pseudo-experiments. In
each experiment, we generate the \mjj pseudo-data for the SM processes,
taking into account the correlations between the yields, and then perform a fit
to each pseudo-data \mjj distribution.
The results for both the muon and electron channels
indicate that there is a small bias (-8.6\% and -6.6\%) in the
WW+WZ yield, corresponding to less than 0.4 standard deviations, and that the
fit slightly overestimates the uncertainty on the yield.
These effects are corrected for in the final result.
The validation procedure shows that biases in all
background yields and errors are small.
The fit results for the background components are
statistically consistent with the expectations,
with the exception of W+jets, where 11\% fewer events for muons 
and 15\% fewer events for electrons, compared to the expectation, are observed.
Overall, the approach produces a high quality model of the data (Fig.~\ref{fig:Fig1}(a)), where the pull distribution is consistent with 0 (Fig.~\ref{fig:Fig1}(c)), and allows us to extract the diboson peak (Fig.~\ref{fig:Fig1}(b)).

Systematic uncertainties arising from the jet energy are estimated from
W bosons decaying hadronically in a sample of semileptonic
\ttbar\ events. The mean and resolution of the reconstructed dijet
mass distribution in data agree to within 0.6\% of
the expectations from simulation (this discrepancy is 
accounted for as an explicit systematic uncertainty), with negligible effect on acceptance. 
A small difference in \MET
resolution~\cite{Chatrchyan:2011tn} between data and simulation
affects the signal acceptance at the 0.5\% level.
Further systematic uncertainties on the signal yield
are due to the uncertainty on the trigger efficiency in
data (1\%), and on the lepton reconstruction and
selection efficiencies (2\%)~\cite{VBTF}.
The uncertainty due to the b-jet veto is negligible.
The uncertainty
in the luminosity measurement is 2.2\%~\cite{lumiPAS}.
The uncertainty in acceptance arising from theoretical uncertainties (evaluated using \MADGRAPH and \MCFM samples),
including parton distribution functions and additional jet rejection, is 4\%.

As listed in Table~\ref{tab:yields}, we
observe $2700 \pm 340 \stat \pm 360\syst$ WW+WZ events,
in agreement with the SM expectation.
This result corresponds to a significance of 8.8 standard deviations
when computed using a simple likelihood
ratio~\cite{Junk:1999kv}, where the background yield uncertainties 
(Table~\ref{tab:yields}) and errors on $\alpha$, $\beta$ (Eq.~\ref{eqWpjetsShape}) 
are fixed to their fitted values.
Using the profile likelihood ratio~\cite{Junk:1999kv},
where these parameters are allowed to vary, the significance
becomes 4.3 standard deviations.
We compute the WW+WZ total cross section as
$\sigma = N_{\text{Sig}}/ (\mathcal{A} \, \varepsilon \, {\mathcal{L}})$,
where $N_{\text{Sig}}$ is the number of extracted signal events,
$\mathcal{A}$ is the signal acceptance corrected for the branching fractions,
$\varepsilon$ is the overall efficiency for event selection,
and ${\mathcal{L}}$ is the integrated luminosity.
In the acceptance calculation we assume the SM value for the WW to WZ production ratio.
The values of $N_{\text{Sig}}$ and $\mathcal{A}\varepsilon$
are given in Table~\ref{tab:yields} separately for the muon and electron channels.
Combining the results from the muon and electron channels, we obtain
$\sigma(\Pp\Pp\to\text{WW}+\text{WZ}) = 68.9 \pm 8.7\stat\pm 9.7\syst \pm 1.5\lum\unit{pb}$,
which is in agreement with the NLO prediction~\cite{Campbell:2011bn}, $65.6 \pm 2.2\unit{pb}$,
that includes the contribution from $\cPg\cPg\to\text{WW}$.
The total cross section computed in the muon channel, $73.8 \pm 15.1\unit{pb}$,
is consistent with that obtained in the electron channel, $60.8 \pm 21.5\unit{pb}$,
where the statistical and systematic uncertainties have been combined in quadrature.
\begin{figure*}[tbh]
  {\centering
    \includegraphics[width=0.48\textwidth]{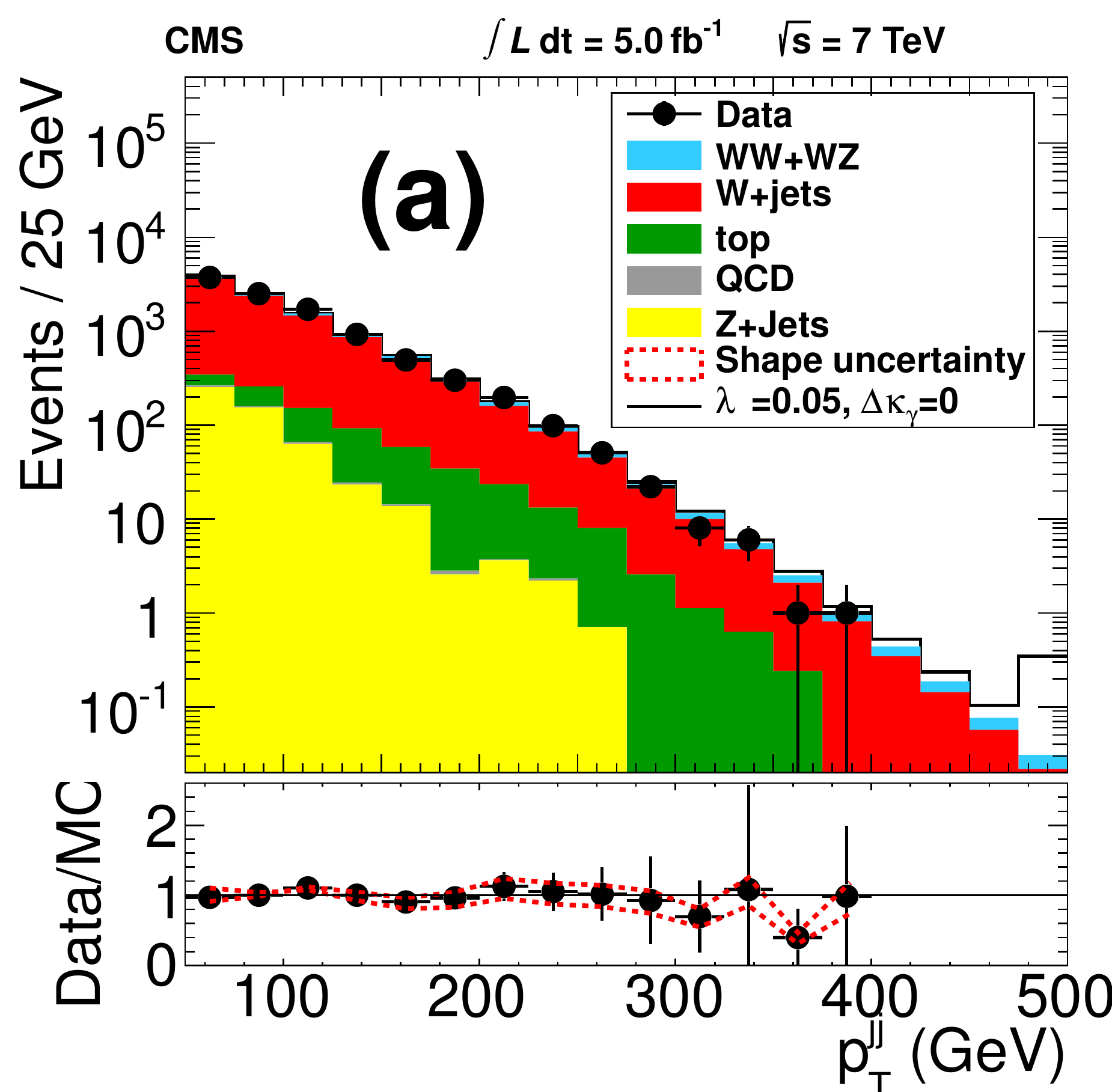}
    \includegraphics[width=0.48\textwidth]{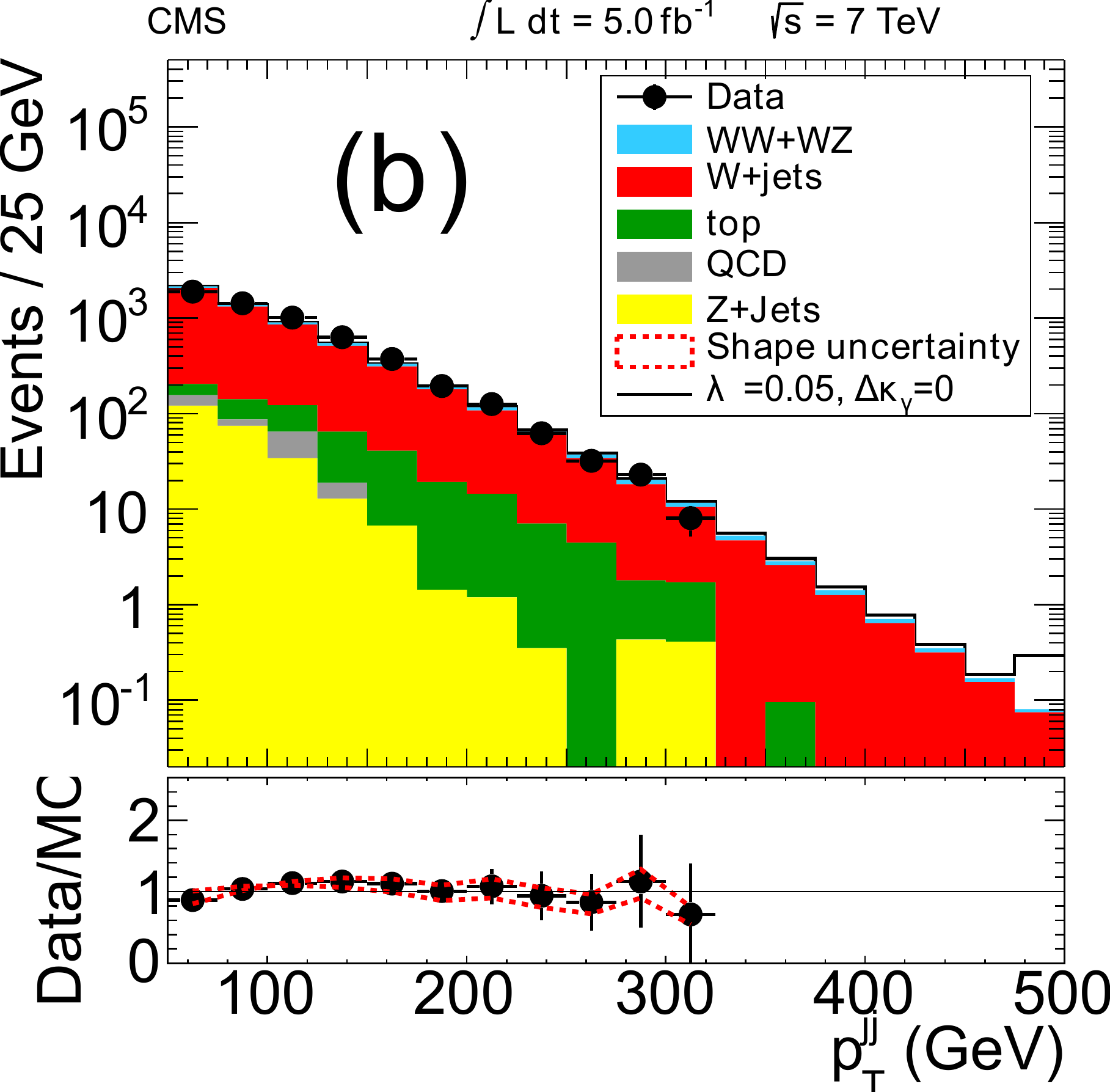}
    \caption{Dijet \pt distributions for (a) the muon and (b) the electron
    channels after full selection and with the requirement
    $75\GeV < \mjj < 95\GeV$.
    The stacked histogram shapes are taken from simulation
    or, where applicable, from data-driven estimates.
    They are normalized according to the fit to the observed
    \mjj spectrum in data. Below we show the Data/MC ratio with the (dashed) red lines corresponding to the shape uncertainty.
    The last bin includes the overflow.
    }
    \label{fig:Fig2}}
\end{figure*}

\begin{figure}[tbh]
  {\centering
    \includegraphics[width=\columnfigure]{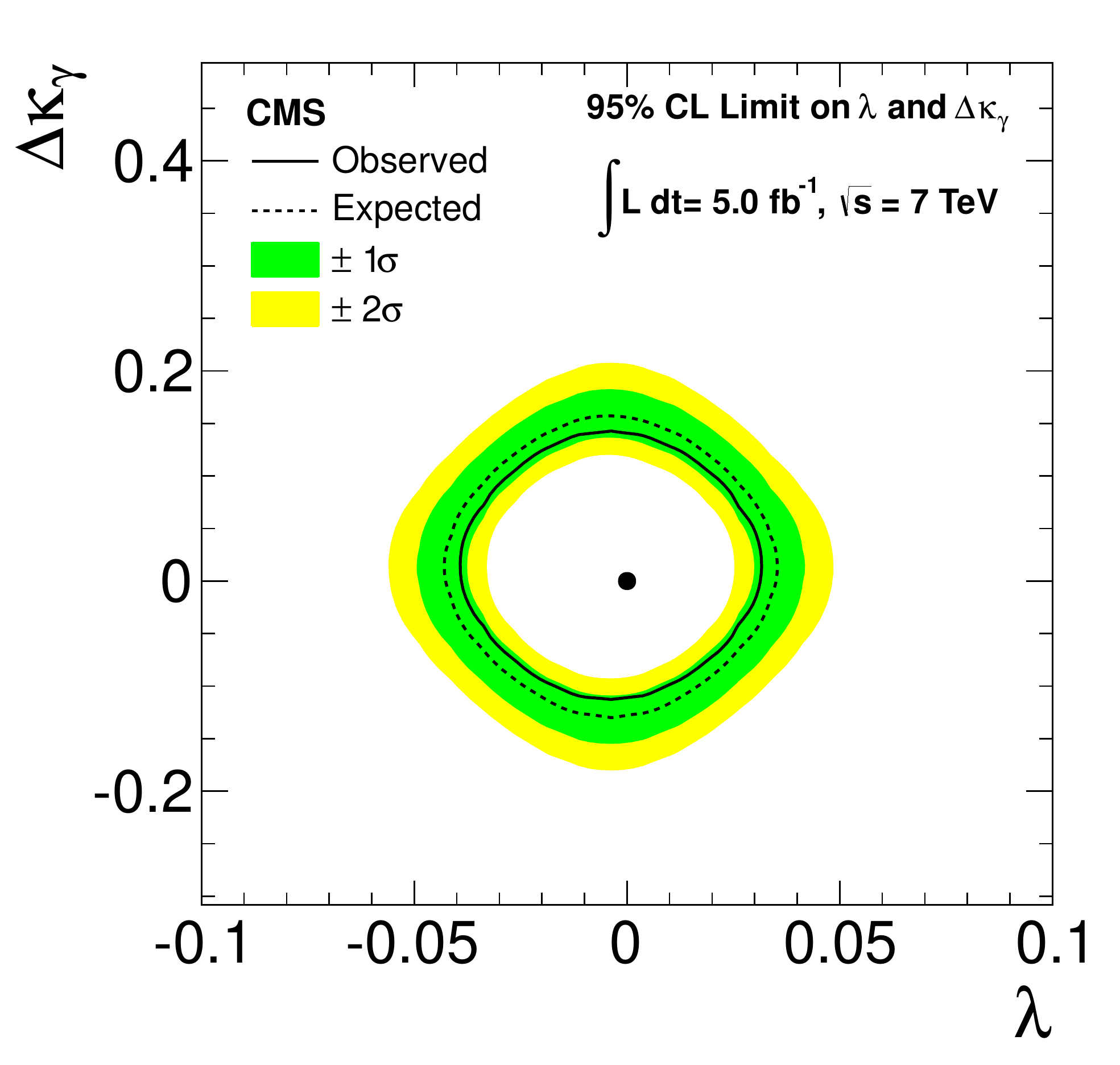}
    \caption{Observed (solid) and expected (dashed) exclusion limits at 95\% CL for
    anomalous triple gauge couplings.
    The dark green and light yellow bands correspond to
    the one and two sigma intervals, respectively, in the
    expected limit distribution.
    The SM expectation is shown by the solid dot.
    }
    \label{fig:Fig3}}
\end{figure}

Measurements of electroweak diboson production can be translated into measurements of
gauge boson self-couplings, 
which are among the most fundamental
aspects of the SM.
At the leading order, only $s$-channel $\cPq\cPaq$ annihilation
diagrams have a three-boson vertex involving WW$\gamma$ and WWZ couplings
in WW production, and WWZ coupling in WZ production.
Physics beyond the SM can modify these couplings, leading to
observable differences in the cross section and the kinematic
distributions~\cite{Dixon:1999di}.
We search for anomalous triple gauge couplings (aTGCs) using an effective
Lagrangian described by the following HISZ (Hagiwara, Ishihara, Szalapski, Zeppenfeld) param\-etrization
without form factors~\cite{PhysRevD.48.2182}:
$\lambda_{\gamma} = \lambda_Z = \lambda$,
$\Delta{\kappa_Z} = \Delta{g_1^Z}-\Delta{\kappa_\gamma} \cdot \tan^2\theta_{\mathrm{W}}$.
We use the dijet \pt distribution (with most of the discriminating power coming from the last bin), 
shown in Fig.~\ref{fig:Fig2},
as the observable after requiring  $75\GeV < \mjj < 95\GeV$ to enhance signal purity.
The dependence of the distribution on specific aTGCs
is modeled by reweighting the \PYTHIA simulation of WW+WZ to
the \MCFM~\cite{MCFM} NLO predictions.
We account for systematic uncertainties arising from luminosity,
signal selection efficiency (via comparisons to \MCFM samples employing alternate 
choices of PDFs as well as factorization and renormalization scales), signal shape,
and from the normalization and shape of the SM processes.
We find no evidence for aTGCs.
Given the tight bound on the parameter $\Delta{g_1^Z}$~\cite{Beringer:1900zz},
we assume the SM value~($\Delta{g_1^Z}=0$) and set limits on
the two parameters $\lambda$ and $\Delta{\kappa_\gamma}$.
Exclusion limits at 95\% confidence level (CL) in the two-dimensional
$\lambda$-$\Delta{\kappa_\gamma}$ plane,
computed using the modified frequentist CL${}_{S}$~\cite{Junk:1999kv,CLS}
technique with profile likelihood as the test statistic,
are shown in Fig.~\ref{fig:Fig3}. The limit setting procedure combines 
fit results from muon and electron channels
We obtain the following one-dimensional observed 95\%
CL limits assuming the SM value for the other parameter:
$ -0.038 < \lambda < 0.030$,
$ -0.11 < \Delta{\kappa_\gamma} < 0.14$.
These limits are competitive with, and in some cases improve upon,
the sensitivity of the combined LEP experiments listed in
Refs.~\cite{Beringer:1900zz,Abdallah:2008sf,Schael:2004tq,Abbiendi:2003mk,Achard:2004ji}.
The ATLAS Collaboration recently reported
limits in the fully leptonic channel for WZ~\cite{ATLAS:2012wz} and WW~\cite{ATLAS:2012ww} production.
Limits obtained from fully leptonic channels are weaker due to the smaller branching fractions.

In summary, a measurement of the sum of the inclusive WW and WZ production cross sections
has been performed using events containing a leptonically decaying W and two jets.
The measured value for the cross section is
$\sigma(\Pp\Pp\to\text{WW}+\text{WZ}) = 68.9 \pm 8.7\stat\pm 9.7\syst  \pm 1.5\lum\unit{pb}$,
which is consistent with the SM prediction. This is the first measurement of WW+WZ production in pp collisions using
this signature. No evidence for anomalous triple gauge couplings is found,
and limits are set on their magnitudes.

{\tolerance=500
We congratulate our colleagues in the CERN accelerator departments for the excellent performance of the LHC and thank the technical and administrative staffs at CERN and at other CMS institutes for their contributions to the success of the CMS effort. In addition, we gratefully acknowledge the computing centers and personnel of the Worldwide LHC Computing Grid for delivering so effectively the computing infrastructure essential to our analyses.  Finally, we acknowledge the enduring support for the construction and operation of the LHC and the CMS detector provided by the following funding agencies: BMWF and FWF (Austria); FNRS and FWO (Belgium); CNPq, CAPES, FAPERJ, and FAPESP (Brazil); MES (Bulgaria); CERN; CAS, MoST, and NSFC (China); COLCIENCIAS (Colombia); MSES (Croatia); RPF (Cyprus); MoER, SF0690030s09 and ERDF (Estonia); Academy of Finland, MEC, and HIP (Finland); CEA and CNRS/IN2P3 (France); BMBF, DFG, and HGF (Germany); GSRT (Greece); OTKA and NKTH (Hungary); DAE and DST (India); IPM (Iran); SFI (Ireland); INFN (Italy); NRF and WCU (Korea); LAS (Lithuania); CINVESTAV, CONACYT, SEP, and UASLP-FAI (Mexico); MSI (New Zealand); PAEC (Pakistan); MSHE and NSC (Poland); FCT (Portugal); JINR (Armenia, Belarus, Georgia, Ukraine, Uzbekistan); MON, RosAtom, RAS and RFBR (Russia); MSTD (Serbia); SEIDI and CPAN (Spain); Swiss Funding Agencies (Switzerland); NSC (Taipei); TUBITAK and TAEK (Turkey); STFC (United Kingdom); DOE and NSF (USA).
\par}

\vspace*{6ex} 
\bibliography{auto_generated}   

\cleardoublepage \appendix\section{The CMS Collaboration \label{app:collab}}\begin{sloppypar}\hyphenpenalty=5000\widowpenalty=500\clubpenalty=5000\textbf{Yerevan Physics Institute,  Yerevan,  Armenia}\\*[0pt]
S.~Chatrchyan, V.~Khachatryan, A.M.~Sirunyan, A.~Tumasyan
\vskip\cmsinstskip
\textbf{Institut f\"{u}r Hochenergiephysik der OeAW,  Wien,  Austria}\\*[0pt]
W.~Adam, E.~Aguilo, T.~Bergauer, M.~Dragicevic, J.~Er\"{o}, C.~Fabjan\cmsAuthorMark{1}, M.~Friedl, R.~Fr\"{u}hwirth\cmsAuthorMark{1}, V.M.~Ghete, J.~Hammer, N.~H\"{o}rmann, J.~Hrubec, M.~Jeitler\cmsAuthorMark{1}, W.~Kiesenhofer, V.~Kn\"{u}nz, M.~Krammer\cmsAuthorMark{1}, I.~Kr\"{a}tschmer, D.~Liko, I.~Mikulec, M.~Pernicka$^{\textrm{\dag}}$, B.~Rahbaran, C.~Rohringer, H.~Rohringer, R.~Sch\"{o}fbeck, J.~Strauss, A.~Taurok, W.~Waltenberger, G.~Walzel, C.-E.~Wulz\cmsAuthorMark{1}
% ...
\vskip\cmsinstskip
\dag:~Deceased\\
1:~~Also at Vienna University of Technology, Vienna, Austria\\
% ...

\end{sloppypar}
\end{document}